\documentclass[runningheads]{llncs}
\usepackage{graphicx}
\usepackage{booktabs,makecell,tabularx}
\usepackage{todonotes}
\usepackage{amssymb}
\usepackage{subfiles}
\renewcommand{\todo}[1]{{\color{red} TODO: {#1}}}
\begin{document}
\title{Confident Head Circumference Measurement from Ultrasound with Real-time Feedback for Sonographers
}

%
\titlerunning{Confident HC Measurement for fetal US}
%
\author{Samuel Budd\inst{1}, 
Matthew Sinclair\inst{1},
Bishesh Khanal\inst{3}\inst{2}, 
Jacqueline Matthew\inst{2}, 
David Lloyd\inst{2}, 
Alberto Gomez\inst{2}, 
Nicolas Toussaint\inst{2}, 
Emma Robinson\inst{2}, 
Bernhard Kainz\inst{1}}
%
\authorrunning{S Budd et al.}
%
\institute{Imperial College London, Dept. Computing, BioMedIA, London, UK  \and
King's College London, ISBE, London, UK \and NAAMII, Kathmandu, Nepal \\ \email{samuel.budd13@imperial.ac.uk}
}
\maketitle              

\begin{abstract}
Manual estimation of fetal Head Circumference (HC) from Ultrasound (US) is a key biometric for monitoring the healthy development of fetuses. Unfortunately, such measurements are subject to large inter-observer variability, resulting in low early-detection rates of fetal abnormalities. To address this issue, we propose a novel probabilistic Deep Learning approach for real-time automated estimation of fetal HC. This system feeds back statistics on measurement robustness to inform users how confident a deep neural network is in evaluating suitable views acquired during free-hand ultrasound examination. In real-time scenarios, this approach may be exploited to guide operators to scan planes that are as close as possible to the underlying distribution of training images, for the purpose of improving inter-operator consistency. 
We train on freehand ultrasound data from over 2000 subjects (2848 training/540 test) and show that our method is able to predict HC measurements within $1.81 \pm 1.65mm$ deviation from the ground truth, with 50\% of the test images fully contained within the predicted confidence margins, and an average of $1.82 \pm 1.78 mm$ deviation from the margin for the remaining cases that are not fully contained.
\end{abstract}
\section{Introduction}

Fetal Ultrasound (US) scanning is a vital part of ensuring good health of mothers and fetuses during and after pregnancy. Accurate anomaly detection and assessment of fetal development from US scans are required to ensure that the best care is given at the earliest identifiable stage. In many countries a mid-trimester US scan is carried out between 18-22 weeks gestation as a part of standard prenatal care. `Standardized plane' views are used to acquire images in which distinct anatomical features can be extracted~\cite{Fasp2018}. From some of these standard plane views, measurements of the head, abdomen and femur are most commonly used to predict fetal age and weight, and are the key biometrics identified from US. Biometrics acquired longitudinally can be used to predict the fetal development trajectory. Unfortunately, rates for early detection of fetal abnormalities are low, largely due to the high level of skill required by the sonographer to perform such scans and extract the relevant biometrics~\cite{Sarris2012}. 

Recently, automatic US scanning approaches have been developed using deep learning~\cite{Baumgartner2017SonoNet:Ultrasound}, which mitigate the problems of manual US measurement through automatic detection of diagnostically relevant anatomical planes. Such systems have allowed development of robust automated methods for estimation of anatomical biometrics~\cite{Sinclair2018Human-levelNetworks,Wu2017CascadedSegmentation} in diverse acquisition conditions with various imaging artefacts, outperforming non-deep learning approaches~\cite{Carneiro2008DetectionTree,Li2018AutomaticFitting,Rueda2014EvaluationChallenge}. Critically, such methods only provide point estimates of HC without confidence or uncertainty measures, and do not provide any means to assess the quality of individual measurements during real-time scans. This can lead to many, potentially contradicting, measurements without any means to control the trustworthiness of the predictions during examination or retrospectively. 
To this end, several approaches have been proposed for estimation of uncertainty in Deep Networks. These include Monte-Carlo Dropout (MC Dropout), the most common dropout method which has been shown to model a posterior mixture of Gaussians well. Weights in a deep neural network are `dropped' randomly during inference with a given probability $p$ which has been shown to approximate Bayesian inference in deep Gaussian processes~\cite{Gal2016DropoutGhahramani}. 
In addition, ensemble approaches produce $N$ prediction samples per input image by training a set of $N$ separate networks for the same task. The results are then combined to produce a final segmentation which seems to offer a good trade-off between robustness and accuracy~\cite{KamnitsasEnsemblesSegmentation}. Finally, the Probabilistic U-Net represents a generative segmentation model based on a combination of a U-Net with a conditional variational autoencoder. This is capable of producing an unlimited number of plausible hypotheses, reproducing the possible segmentation variants as well as the frequencies with which they occur~\cite{Kohl2018}. \\

\noindent\textbf{Contribution:}
In this paper, we extend upon a state-of-the-art convolutional Deep Learning approach for automatic fetal HC measurement \cite{Sinclair2018Human-levelNetworks} to develop a new approach for automated probabilistic fetal HC with real-time feedback on measurement robustness. Two probabilistic deep learning methods are evaluated: MC Dropout during inference and Probabilistic U-Net. These are used to return an ensemble of segmentations, from which upper and lower bounds on the measurement are generated. In addition, we propose the derivation of a `variance score', used to reject acquired images that produce sub-optimal HC measurements. In this way, the system will guide operators towards acquiring optimal US views, resulting in more consistent and accurate measurements.

\section{Method}
\begin{figure}[htb]
    \makebox[\textwidth][c]{\includegraphics[width=\textwidth]{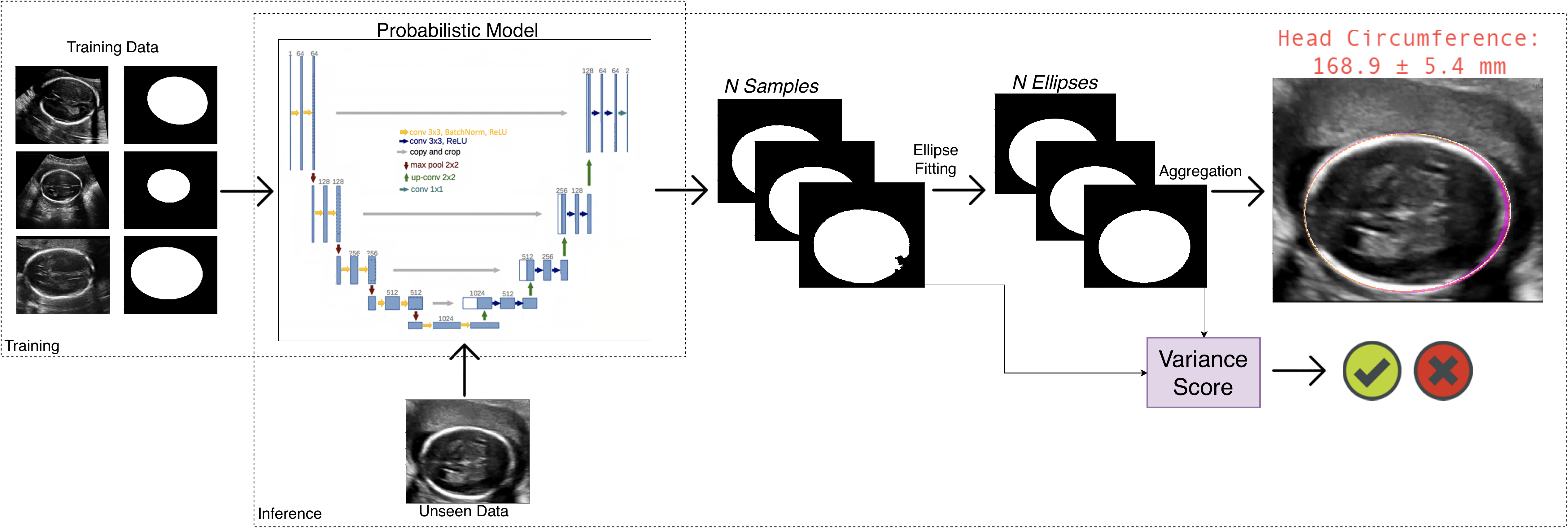}}%
    \caption{Overview of our proposed method. We train a probabilistic model using the available training data. During inference we take $N$ samples from our model, fit ellipses to each sample and aggregate these ellipses to extract a HC value and an upper and lower bound on that HC value. Various outputs of the pipeline are used to calculate different variance scores given a set of $N$ samples. As a proof of concept we extract a threshold such that test cases whose variance score is outside the threshold are rejected, and inside are accepted.}
    \label{fig:overview}
\end{figure}

\noindent\textbf{Biometric estimation:}
Our HC estimation builds on the approach developed in \cite{Sinclair2018Human-levelNetworks} which achieves human level performance. First, a U-Net \cite{RonnebergerU-Net:Segmentation} segmentation network masks out the head from an US image. Then, an ellipse is fitted to the segmented contours~\cite{Fitzgibbon1996DirectEllipses} from which the ellipse parameters can be obtained in mm. We extract ellipse centroid co-ordinates ($c_x$ and $c_y$), major and minor axis radii ($a$ and $b$) each in pixels, and the angle of rotation ($\alpha$) and estimate HC using the Ramanujan approximation II \cite{Barnard2001InequalitiesEllipse} as 
$
    HC = \pi (a + b) (1 + \frac{3h}{10 + \sqrt{4 - 3h}})s_{xy}
$
where
$
        h = \frac{(a-b)^2}{(a+b)^2}
$.
The error of this approximation is $O(h^{10})$ which for more circular ellipses is negligible. This ellipse fitting process mimics the sonographer's manual actions when extracting a HC measurement during fetal US screening.\\

\noindent\textbf{Probabilistic segmentation:}
Given the inherent variability between sonographers' annotations in the training data, we generate a set of $N$ plausible segmentations from a single input using the following methods: 

\emph{i)} \textit{MC Dropout}: We randomly drop weights of the network with probability $p$ to predict $N$ segmentation samples. Here, single-sample experiments ($N=1$) were used to optimise the configuration of the network. This led to implementation of a single dropout layer ($p=0.6$) before the bottleneck layer of the U-Net during inference. 

\emph{ii)} \textit{Probabilistic U-Net}: We sample a set of $N$ plausible segmentations using this method~\cite{Kohl2018} where we follow the same training scheme as \cite{Kohl2018}.\\


\noindent\textbf{Variance Estimation:}
With a probabilistic mapping function $g_P(X) = \hat{X}_i$, in our case a deep probabilistic neural network, we can map a continuous input image to a possible segmentation mask $\hat{X}_i$.
We assume a deterministic function $f(\hat{X}_i) = [a, b, \theta, x_c, y_c ]^T$, with semi-major axis length $a$, semi-minor axis length $b$, angle of
orientation  $\theta$ and center $C(x_c,y_c)$, which provides a least square solution to the ellipse fitting problem to the set of points $\hat{X}$ as proposed by \cite{PRASAD20131449}. Based on $f(\hat{X}_i)$ we can evaluate hypotheses for their suitability to act as a metric to measure robustness during inference given $N$ prediction samples from $g_P(X)$. These proposed metrics are 

\emph{h1)} \textit{Ellipse parameter variance}: $ \sum^5_i(\mathrm{Var}(f(\hat{X}_n)_i))$; 

\emph{h2)} \textit{Total ring area}: $\sum(f(\bigcup_{i=1}^{N} \hat{X}_i) - f(\bigcap_{i=1}^{N} \hat{X}_i)) \cdot s_{xyz}$, where $s_{xyz}$ scales $\hat{X}_i$ to world space in $mm$; 

\emph{h3)} \textit{Mask classification entropy}:  $\sum_{x,y}^K \underline{\hat{X}}(x,y)\log(\underline{\hat{X}}(x,y))$, where $K$ is the number of pixels in $\underline{\hat{X}} \in \mathbb{Z}_2 $ after $argmax(\hat{X}_i)$ class assignment and $\underline{\hat{X}} = \frac{1}{N}\cdot\sum^N_i\hat{X}_i $; and 

\emph{h4)} \textit{Softmax confidence entropy}: given $\hat{X}_i \in \mathbb{R} $ before class assignment, after conversion of the network's final layer's logits with  $Softmax(x_i) = \frac{\exp(x_i)}{\sum^i\exp(x_i)}$, the resulting $\hat{X}_i^\ast$ can be interpreted as two-element prediction confidence $[p_f, p_b]_i = \hat{X}_i^\ast(x,y)$ for foreground $p_f$ and background $p_b$. Thus we can estimate class-agnostic prediction entropy by $\sum_i^K p_i \log(p_i)$  where $p_i =\sum_i^N \max([p_f, p_b]_i)$.

\section{Experiments and Results}

\noindent\textbf{Data:}
Our base dataset, named subsequently as Dataset A, consists of 2,724 two-dimensional US examinations from volunteers 
at 18-22 weeks gestation, acquired and labelled during routine screening by 45 expert sonographers. Several images were taken during each session, including the standard transverse brain view at the posterior horn of the ventricle (TV) plane used for HC measurement. 
This data was combined with the HC18 Challenge~\cite{vandenHeuvel2018AutomatedImages} dataset which consists of 1334 two-dimensional US images of the standard plane that is used to measure HC, each image is 800x540 pixels with a pixel size ranging from 0.052mm to 0.326mm. Each image in the training set has an accompanying manual annotation of the HC (ellipse outline) performed by a single trained sonographer~\cite{vandenHeuvel2018AutomatedImages}. We resample all images to $320 \times 384$ pixels, and produce a head mask from the expert ground truth delineation. Training data is randomly flipped both horizontally and vertically, and a random rotation ($\pm 5^\circ$)is performed. \\



\noindent\textbf{Single-Sampling Experiments:} In the first instance, single-sample experiments, generating a single segmentation and HC measurement ($N=1$) per subject, were used to verify the performance of the proposed model against the state-of-the-art~\cite{Sinclair2018Human-levelNetworks}. Table \ref{res:det} reports performance measures for all \textit{single-sampling} experiments. These show comparable performance relative to \cite{Sinclair2018Human-levelNetworks} for our U-Net implementation, trained on Dataset A. This result improves further when the same model is trained on Dataset A and HC18 data. MC dropout during training further improves the result. For subsequent analysis, all experiments for MC Dropout (during inference) use the combined data and are trained using MC dropout. \\
\begin{table}[h!]
\caption{Single sample results of three U-Net's. \textbf{Baseline}: Trained on Dataset A data only. \textbf{Dataset A + HC18}: Trained on Dataset A data and HC18 Challenge data transformed to same format as Dataset A data. \textbf{Dropout}: Trained on Dataset A and HC18 Challenge data with dropout ($p=0.6$ value found to be best performing in variety of dropout configurations). We compare the Mean absolute difference between the final HC measurement, the DICE overlap of the fitted ellipse with the ground truth ellipse, and the Hausdorff distance between the outline of the fitted ellipse and the outline of the ground truth ellipse. Results calculated on Dataset A test data.}\label{res:det}
\centering
\begin{tabular}{llll}
\hline
& \multicolumn{1}{p{3cm}}{\centering Mean abs difference \\ $\pm$ std (mm)} & \multicolumn{1}{p{3cm}}{\centering Mean DICE \\ $\pm$ std  (\%)} & \multicolumn{1}{p{3cm}}{\centering Mean Hausdorff distance \\ $\pm$ std (mm)} \\
\hline
Baseline             & \multicolumn{1}{p{3cm}}{\centering 2.09 $\pm$ 1.97 }      &  \multicolumn{1}{p{3cm}}{\centering 0.982 $\pm$ 0.011}         &  \multicolumn{1}{p{3cm}}{\centering 1.289 $\pm$ 0.880}           \\
Dataset A + HC18         &  \multicolumn{1}{p{3cm}}{\centering 1.90 $\pm$ 1.90 }             &  \multicolumn{1}{p{3cm}}{\centering 0.982 $\pm$ 0.010 }      & \multicolumn{1}{p{3cm}}{\centering 1.292 $\pm$ 0.791 }           \\
\textbf{Dropout $p=0.6$}   & \multicolumn{1}{p{3cm}}{\centering \textbf{1.808 $\pm$ 1.65}} & \multicolumn{1}{p{3cm}}{\centering\textbf{ 0.982 $\pm$ 0.008}} & \multicolumn{1}{p{3cm}}{\centering \textbf{1.295 $\pm$ 0.664}} \\
\hline
\end{tabular}
\end{table}

\noindent\textbf{Multi-Sampling Experiments:}
 MC Dropout during inference has been compared against a Probabilistic U-Net. Here, multiple ($N$) segmentation predictions are made for each US image. From these, the mean and median of the set of fitted ellipse parameters are used to obtain a single HC value for each test case, and the set of $N$ segmentations are used to obtain an upper and lower bound. Table \ref{res:prob} shows the performance measures for our \textit{multi-sampling} experiments. Results show that we lose performance through aggregating multiple results using the mean or median, although this is likely due to dropout not being applied during inference for single sample experiments. However, the \textit{multi-sampling} methods do allow us to produce an upper and lower bound on the HC value, with an average difference of $1.82 \pm 1.78 mm$ between upper-lower bounds and ground truth HC measurement ($N=10$ samples), for cases where the ground truth is not within the upper-lower bounds (\textit{MC(inf.)}). \\

\begin{table}[h!]
\caption{Multi-Sampling results for the two methods. We report the performance measures of a single-sampled point-predictor (\textit{Det. (Deterministic)}), mean/median of $N=10$ samples from the Probabilistic U-Net (\textit{Prob. U-Net (Probabilistic U-Net)}), and our previous best U-Net with Monte-Carlo dropout during inference (\textit{MC(inf.) (Monte Carlo dropout during inference)}, $p=0.6$). We report the \% ground truth HC values that lie in the calculated upper/lower bound range. This percentage varies significantly with $N$, for \textit{MC(inf.)}: $N=2$: 14.8\%; $N=1000$: 50.4\%. See Supplementary Material Figures 1-3.}\label{res:prob}
\centering
\begin{tabular}{lllll} 
\hline
& \multicolumn{1}{p{2.5cm}}{\centering Mean abs difference \\ $\pm$ std (mm)} & \multicolumn{1}{p{3cm}}{\centering Mean DICE \\ $\pm$ std  (\%)} & \multicolumn{1}{p{3cm}}{\centering Mean Hausdorff distance \\ $\pm$ std (mm)} & \multicolumn{1}{p{2cm}}{\centering $LB \leq HC_{gt} \leq UB (\%)$} \\
\hline
\textit{Det.}  \\
\textbf{MC $p=0.6$}   & \multicolumn{1}{p{2.5cm}}{\centering \textbf{1.81 $\pm$ 1.65}} & \multicolumn{1}{p{3cm}}{\centering\textbf{ 0.982 $\pm$ 0.008}} & \multicolumn{1}{p{3cm}}{\centering \textbf{1.295 $\pm$ 0.664}} & \multicolumn{1}{p{2cm}}{\centering N/A} \\
\hline
\textit{Prob. UNet}    \\
Mean   & \multicolumn{1}{p{2.5cm}}{\centering 2.22 $\pm$ 2.15} & \multicolumn{1}{p{3cm}}{\centering 0.980 $\pm$ 0.011} & \multicolumn{1}{p{3cm}}{\centering 1.413 $\pm$ 0.751} & \multicolumn{1}{p{2cm}}{\centering 20.4} \\
Median   & \multicolumn{1}{p{2.5cm}}{\centering 2.21 $\pm$ 2.15} & \multicolumn{1}{p{3cm}}{\centering 0.980 $\pm$ 0.011} & \multicolumn{1}{p{3cm}}{\centering 1.410 $\pm$ 0.748} & \multicolumn{1}{p{2cm}}{\centering 20.4} \\
\hline
\textit{MC(inf.)} \\
Mean  & \multicolumn{1}{p{2.5cm}}{\centering 2.15 $\pm$ 2.09} & \multicolumn{1}{p{3cm}}{\centering 0.981 $\pm$ 0.010} & \multicolumn{1}{p{3cm}}{\centering 1.313 $\pm$ 0.613} & \multicolumn{1}{p{2cm}}{\centering 27.8}\\
\textbf{Median}   & \multicolumn{1}{p{2.5cm}}{\centering \textbf{2.15} $\pm$ \textbf{2.07}} & \multicolumn{1}{p{3cm}}{\centering \textbf{0.981} $\pm$ \textbf{0.010}} & \multicolumn{1}{p{3cm}}{\centering \textbf{1.307} $\pm$ \textbf{0.604}} & \multicolumn{1}{p{2cm}}{\centering\textbf{27.8}}\\
\hline
\end{tabular}
\end{table}



\begin{figure}
    \makebox[\textwidth][c]{\includegraphics[width=\textwidth]{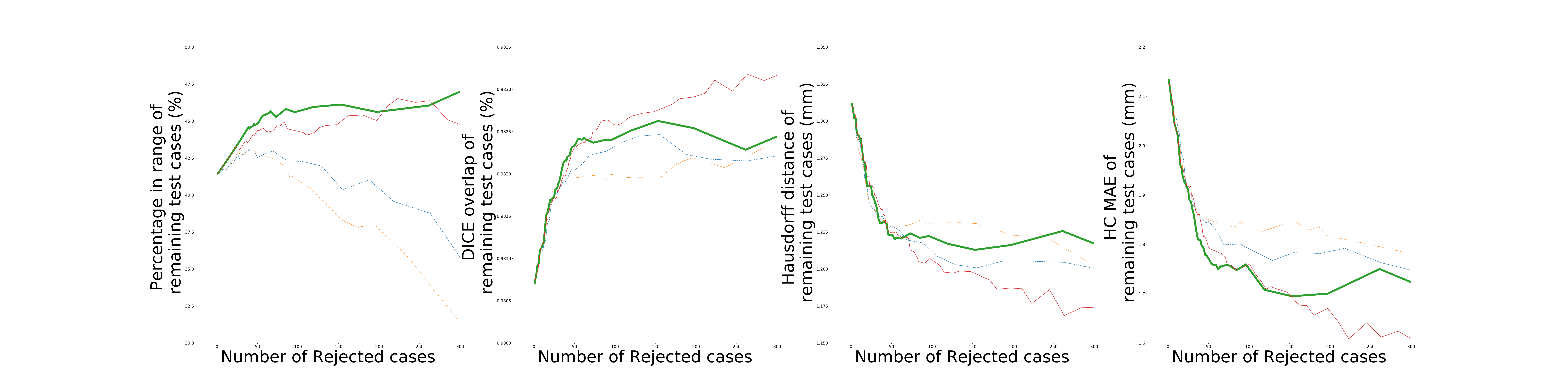}}%
    \vspace{0.1mm}
    \makebox[\textwidth][c]{\includegraphics[width=\textwidth]{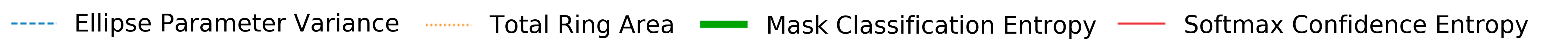}}
    \caption{Plots showing performance measures against the number of rejected test cases. Each measure shows improvement after removing a few test cases for each score (these thresholds vary for each score), however after removing an initial low performing set, the scores power to discriminate between `good' and `bad' images deteriorate. 'Percentage in range' calculated as the percentage of test cases for which the ground truth HC measurement lies within the the predicted upper-lower bounds.}\label{fig:varscore}
\end{figure}

\noindent\textbf{Variance Measure Thresholding:}
Finally, we experiment with each of the variance scores produced over the test set as a means to accept/reject images at test time. We assess their performance by counting the number of accepted/rejected cases for a range of thresholds between zero and one, and how this threshold affects the resulting average performance scores after rejected images are removed from the test set. In this experiment we use only MC dropout during inference ($p=0.6$) which performs best in our previous experiments.
 Figure \ref{fig:varscore} shows graphs depicting how each variance measure can be used to reject test cases, and how rejecting high variance cases can lead to improved performance. In each case we normalise the variance score to lie between 0 and 1, and for each threshold between 0 and 1 we `reject' cases whose variance score is above the threshold. Plots show the performance for remaining `accepted' cases, plotted against the number of `rejected' cases. For most variance scores we obtain an initial performance boost from `rejecting' the worst cases, but after an initial improvement, the variance scores do not delineate `good' from `bad' cases very well. Results suggest that higher measurement variance may indicate sub-optimal imaging plane acquisition.

\begin{figure}
    \makebox[\textwidth][c]{\includegraphics[width=\textwidth]{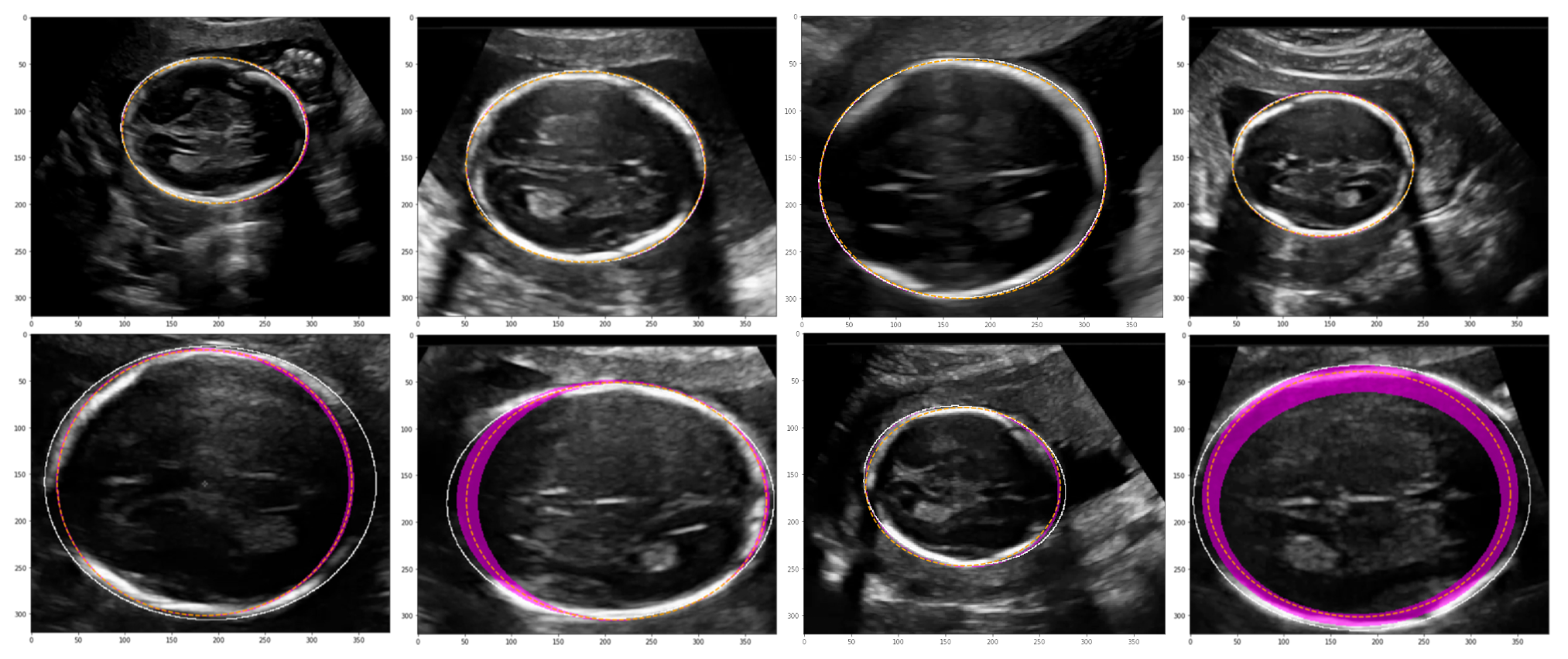}}%
    \caption{Results produced by our model. White line: Ground Truth, Orange dashed line: Mean of sampled ellipse parameters, Pink shaded area: Upper/lower bound range. Top row: High performing images. Bottom row: Low performing images. See Supplementary Material Figures 4 and 5 for more examples and a demo video demonstration.}
    \label{fig:examples}
\end{figure}

\noindent\textbf{Qualitative assessment:}
Figure \ref{fig:examples} shows examples for successful and less model-compliant images using Dropout during inference to produce the samples, where model-compliance captures the proximity of the image to the training data. Note that the best performing examples produce very narrow upper and lower bounds (in this figure where the upper and lower bounds occupy the same pixels the margin is not visible). The worst performing examples show a wider upper and lower bound range but the ground truth ellipse is often not contained within the predicted range. These images often show a lack of clear white presentation of the skull. However, ambiguous segmentation of the regions with missing signal is often reflected in the confidence margin produced, showing greater variation in those image regions, which can be seen clearly in the second example in the bottom row - a wider upper-lower bound area for image regions with low signal from fetal skull. The example on the bottom far right shows missing signal on both sides, which results in a large uncertainty in the ellipses globally due to the compounded effect of missing signal on both sides of the skull.

\section{Discussion}
While we cannot claim our proposed 'variance scores' represent model uncertainty directly, they show some capability to `reject' particularly low performing test cases. In this way, the `variance scores' can be  described as a measurement for the proximity to the variance of the training data of an unseen test sample, which is also desirable, showing the confidence of the network with respect to its capacity and seen training examples. Scenarios in which an operator is present stand to benefit practically using methods introduced in this work, prompting operators to reject sub-optimal measurements by providing real-time feedback during acquisition, thus improving inter-operator consistency. This work lays the foundations for methods by which this can be achieved.

\section{Conclusion}
We demonstrate the effectiveness of probabilistic CNNs to automatically generate HC measurements from US scans, and produce upper-lower bound confidence intervals in real-time. Using multi-sampling probabilistic networks we derive `variance scores', which indicate how confident our network is in generating measurements for a given image. This approach could be used to derive a system which rejects images collected from sub-optimal views, forcing sonographers to take measurements from a view for which the network performs optimally. This could lead to techniques for automated fetal HC measurement, which outperform manual approaches in terms of accuracy and consistency. 

Future directions of this work include exploring alternative methods for multi-sampling networks, alternative segmentation fusion strategies and alternative 'variance scores'. Analysis of new datasets to investigate network bias towards particular datasets is valuable, as well as analysis of cases with anomalous anatomy to verify high performance in the presence of pathologies, clinically the most important cases to identify.

\noindent\textbf{Acknowledgements:} 
This work is supported by the Wellcome Trust IEH 102431, EPSRC (EP/S022104/1, EP/S013687/1), and Nvidia GPU donations.

\bibliographystyle{splncs04}
\bibliography{PhD-iFind}
\newpage
\section*{Supplementary Material}

\subsection*{Example outputs}

\begin{figure}[htb]
    \makebox[\textwidth][c]{\includegraphics[width=\textwidth]{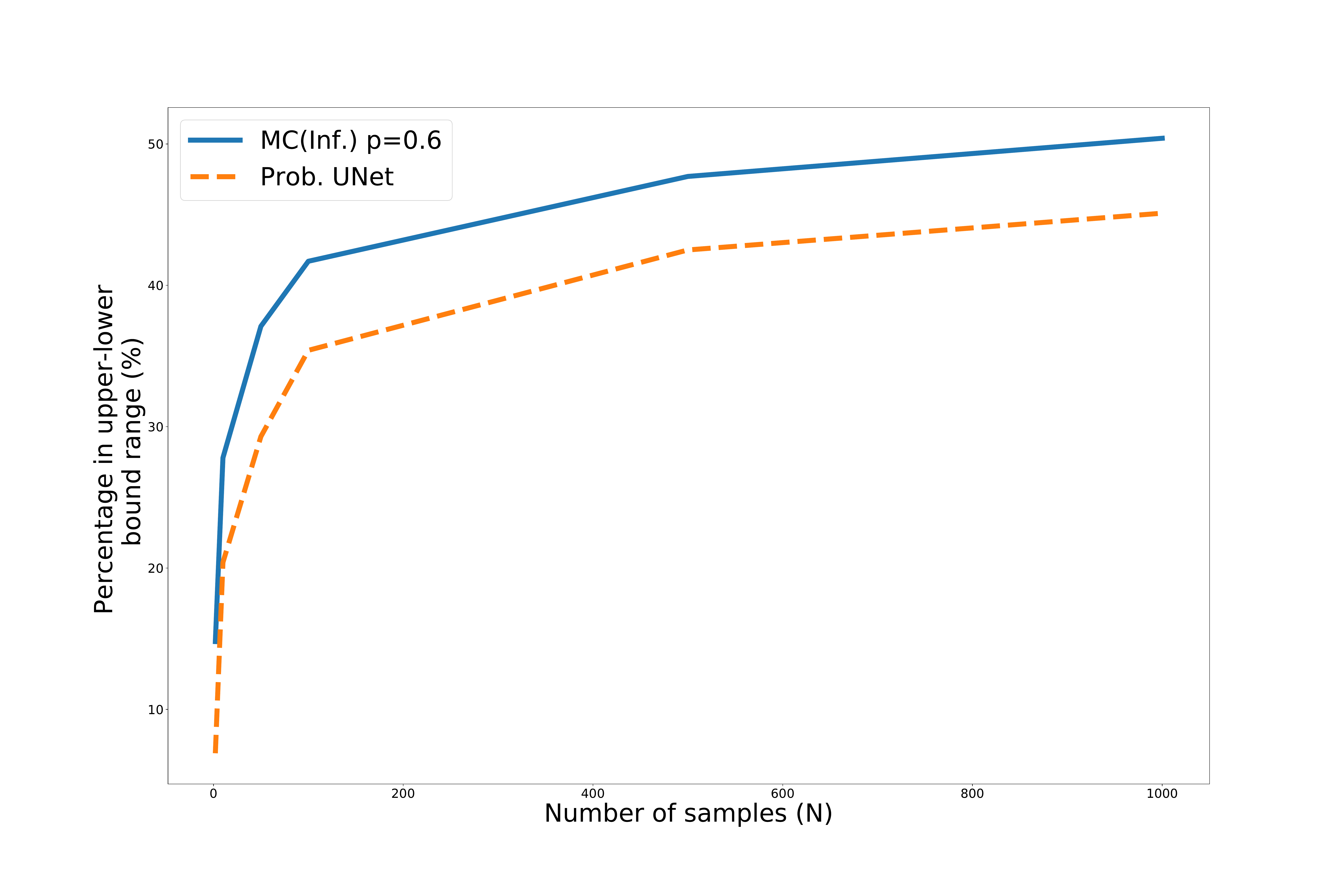}}%
    \caption{Graph showing the effect of taking more samples from the network on the number of ground truth measurements lying between the generated upper and lower bounds. We can see a sharp increase in this percentage as samples increasing, plateauing around 50\% for MC Dropout during inference.}
\end{figure}

\begin{figure}[htb]
    \makebox[\textwidth][c]{\includegraphics[width=\textwidth]{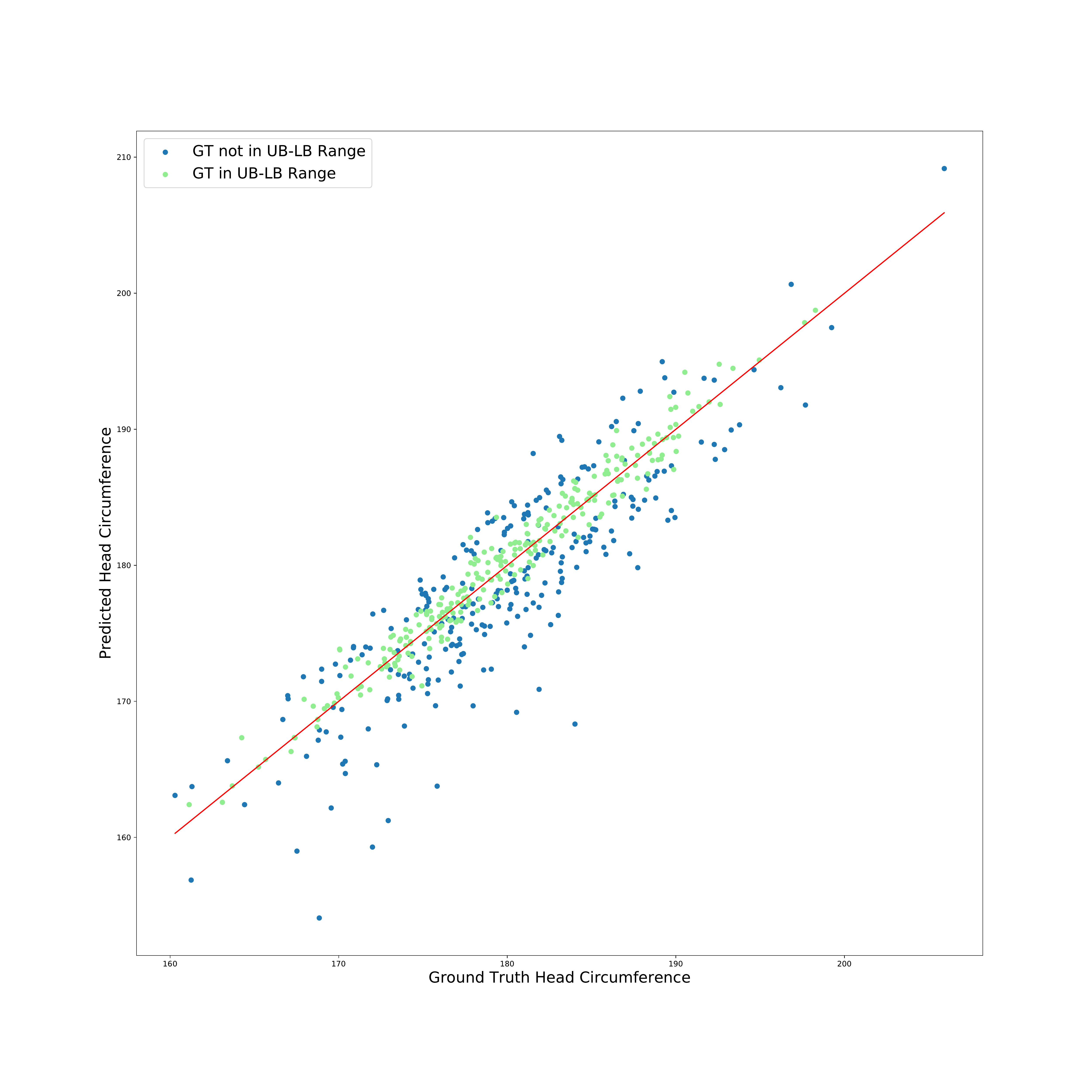}}%
    \caption{Plot showing ground truth HC values against predicted values ($N=1000$), Green dots indicate test cases where the ground truth is contained within the generated upper and lower bounds.}
\end{figure}

\begin{figure}[htb]
    \makebox[\textwidth][c]{\includegraphics[width=\textwidth]{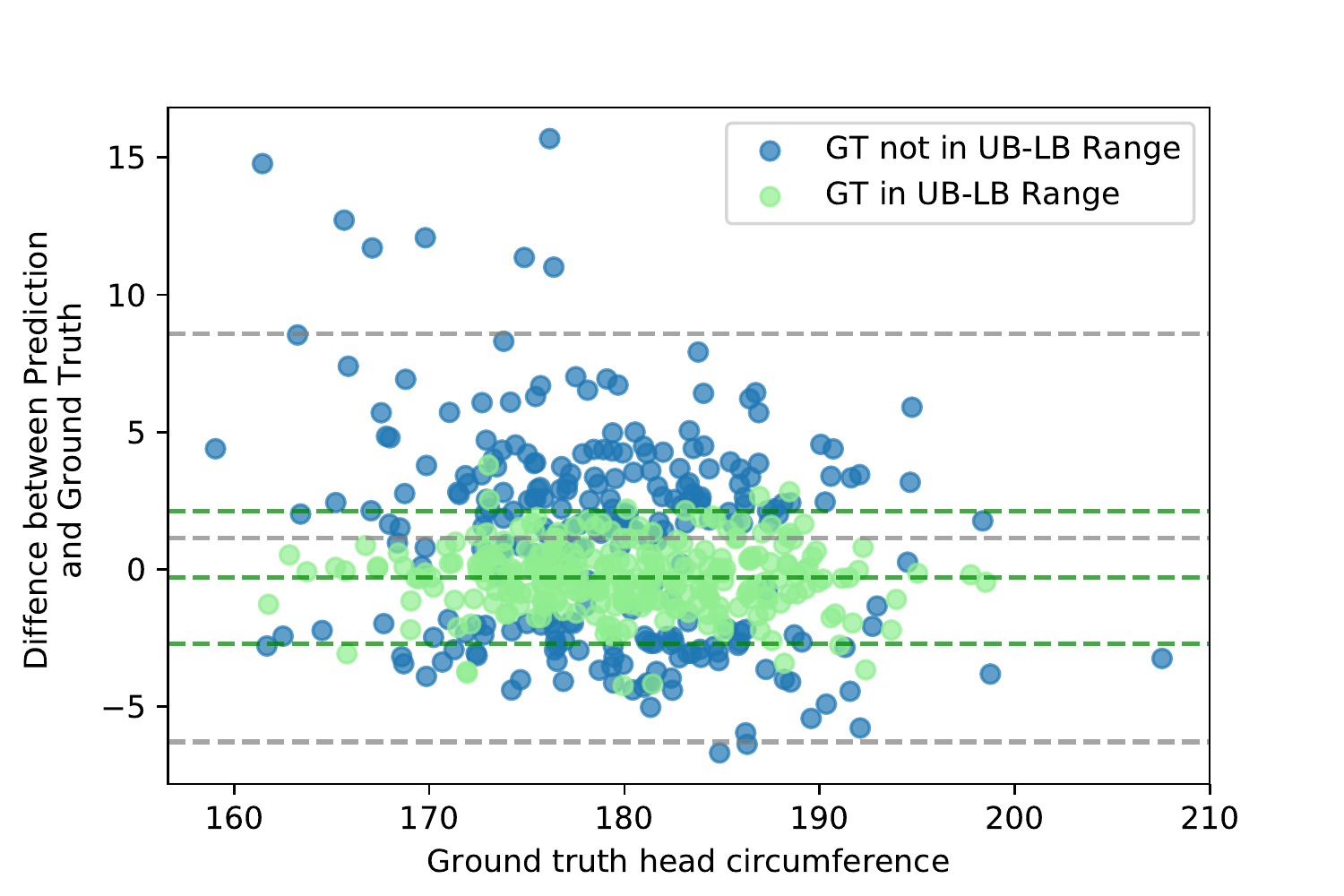}}%
    \caption{Bland-Altman plots for test cases where the ground truth is contained within the upper and lower bounds, and for cases that this is not true.}
\end{figure}

 \begin{figure}[htb]
     \makebox[\textwidth][c]{\includegraphics[width=0.8\textwidth]{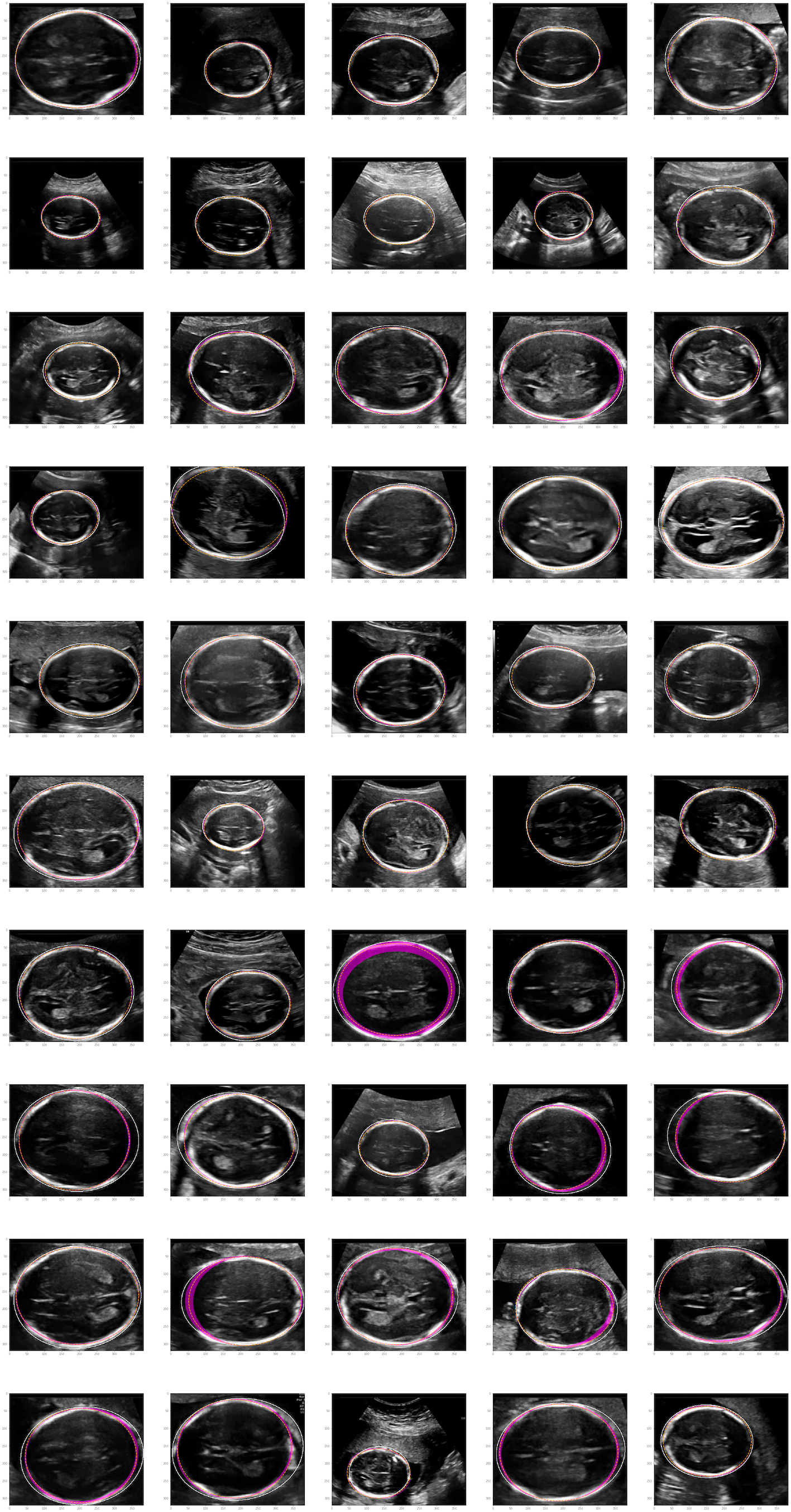}}%
     \caption{Most uncertain test cases}
 \end{figure}

 \begin{figure}[htb]
     \makebox[\textwidth][c]{\includegraphics[width=0.8\textwidth]{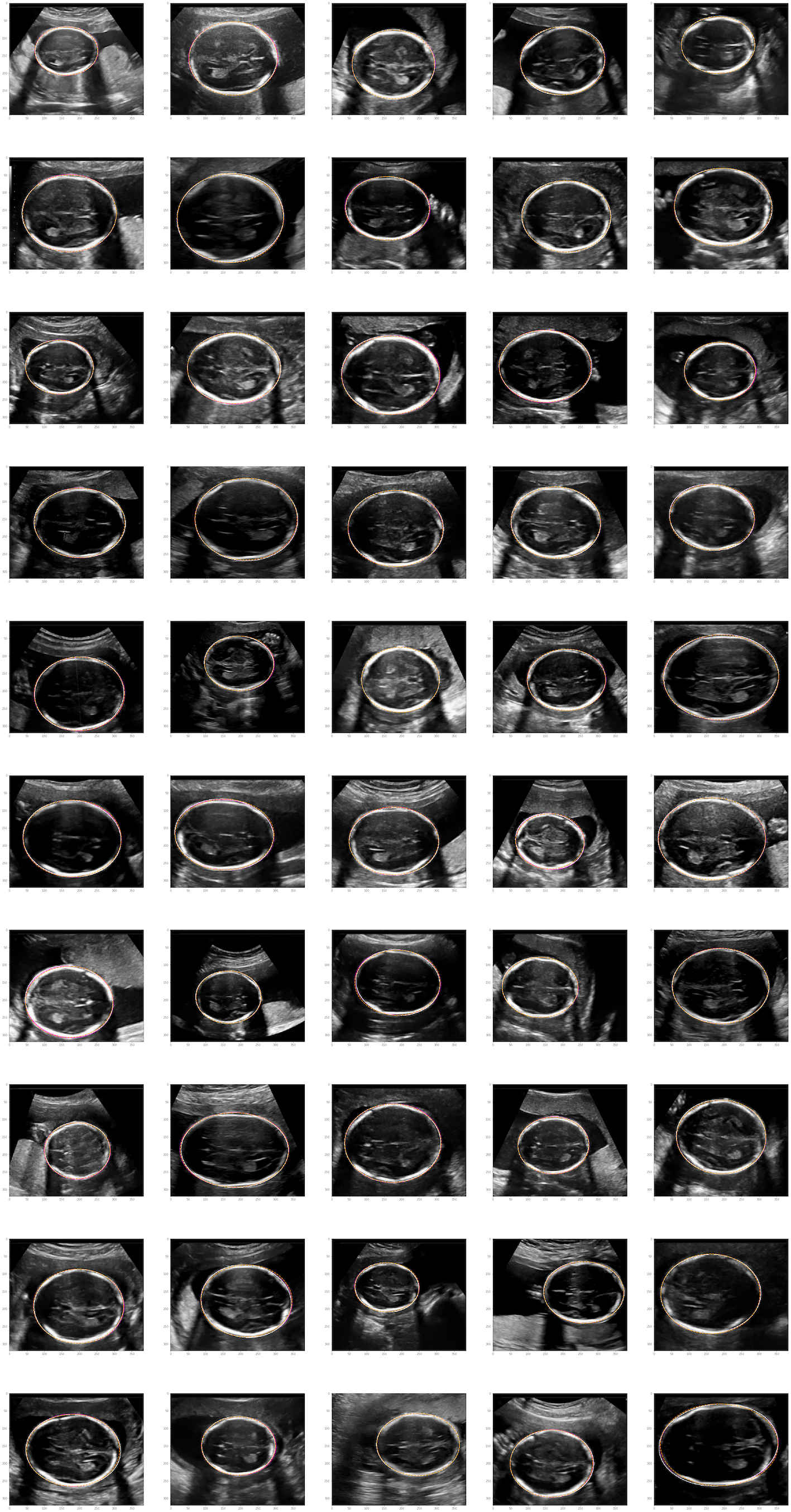}}%
     \caption{Least uncertain test cases}
\end{figure}

\end{document}